\documentclass[prd,showpacs,preprintnumbers,amsmath,amssymb]{revtex4}
\usepackage{graphicx}% Include figure files
\usepackage{dcolumn}% Align table columns on decimal point
\usepackage{bm}% bold math

\def\be{\begin{eqnarray}}
\def\ee{\end{eqnarray}}
\def\bea{\begin{eqnarray}}
\def\eea{\end{eqnarray}}
\def\kT{{\bf k}_\perp}
\def\xT{{\bf x}_\perp}

\def\sT{{\bf s}_\perp}
\def\yT{{\bf y}_\perp}

\def\bT{{\bf b}_\perp}

\def\D2{{\bf \Delta}_\perp^2}
\def\0T{{\bf 0}_\perp}
\begin{document}

%\preprint{draft}

\title{Quark Correlations and Single Spin Asymmetries}% Force line breaks with \\

\author{Matthias Burkardt}
 \affiliation{Department of Physics, New Mexico State University,
Las Cruces, NM 88003-0001, U.S.A.}

\date{\today}% It is always \today, today,


\begin{abstract}
We analyze the Sivers asymmetry in light-cone gauge.
The average transverse momentum of the quark distribution is related 
to the correlation between the quark distribution and the
transverse component of the gauge field at $x^-\pm \infty$.
We then use finiteness conditions for the light-cone Hamiltonian
to relate the transverse gauge field at $x^-=\pm \infty$ to
the color density integrated over $x^-$. This result allows us to 
relate the average transverse momentum of the active quark to
color charge correlations in the transverse plane.
\end{abstract}

\maketitle
\section{Introduction}
Many high energy inclusive hadron processes show surprisingly large
transverse polarizations or single-spin asymmetries \cite{lambda}. 
The most 
prominent example is inclusive hyperon production, but similar
phenomena are observed in many other reactions as well.
For example, in the inclusive photo-production of pions on a
transversely (relative to the photon momentum) polarized nucleon
target, a left-right asymmetry of the produced pions is observed.
Theoretically, two mechanisms (which are not exclusive!) have been 
identified as a potential source of the asymmetry: the Sivers and 
the Collins mechanisms. In the Collins mechanism \cite{collinsold}, 
the asymmetry arises
when a transversely polarized quark fragments into pions with
a left-right asymmetry. In contrast, in the Sivers mechanism 
\cite{sivers} the
asymmetry results from an intrinsic transverse momentum asymmetry of 
the quarks in the target nucleon. At first, such an intrinsic 
transverse momentum asymmetry was expected to vanish due to
time-reversal invariance of the strong interaction.
However, more recently it was realized that, even at high energies, 
the final state interactions (FSI) of the struck quark play an 
important role for the single-spin asymmetry \cite{hwang}. 
Formally the FSI can
be taken into account by introducing an appropriate Wilson line gauge 
link along the trajectory of the ejected quark \cite{collins}.
The gauge invariantly defined transverse momentum distributions
with a gauge link along the light-cone to infinity are no longer
required to vanish to to time-reversal invariance and a nonzero
Sivers asymmetry is possible.

However, while the above reasoning explains the existence of the 
Sivers asymmetry it leaves many questions unanswered, for example
what sign and what magnitude should one expect for the asymmetry,
i.e. is it just an obscure small effect or is it large? If there
is a large asymmetry, what does any information about the asymmetry 
teach us about the structure of the nucleon? In fact, it is also
possible that there is no simple connection between the asymmetry
and ground state properties of the nucleon since the asymmetry
hinges on the inclusion of FSI. In this paper an attempt will be
made to make a step towards answering these questions.  

The paper is organized as follows: in Section 2, we review the
definitions of gauge invariant transverse momentum distributions
and the role of the gauge field at $x^-=\pm \infty$ in the 
light-cone gauge. In Section 3, we use finiteness constraints for
light-cone Hamiltonians to derive operator constraints that allow
one to relate the gauge fields at $x^-=\pm \infty$ to
degrees of freedom at finite $x^-$. In Sections 4 (QED) and 5 (QCD)
we use
these operator constraints to relate the average transverse momentum
to charge (color charge) correlations in the transverse plane.

\section{Initial (final) state interactions and transverse spin
asymmetries}
\label{sec:main}
Ref. \cite{collins} explains how final state interactions (FSI) and 
initial state interactions (ISI) allow the existence of 
T-odd parton distribution functions.
Formally the FSI (ISI) can be incorporated into $\kT $
dependent parton distribution functions (PDFs) by 
introducing a gauge string from each quark field
operator to infinity \cite{collins}
\be
q(x,\kT,\sT) &=& \! \int \frac{dy^- d^2\yT }{16\pi^3}
e^{-ixp^+y^-+i\kT \cdot \yT } \label{eq:P}
\left\langle p \left|\bar{q}(y^-,\yT) \gamma^+
\left[y^-,\yT;\infty^-,\yT \right]
\left[\infty^-,\0T;0^-,\0T\right] q(0)\right|p\right\rangle .
\ee
We use light-front (LF) coordinates, which are defined as:
$y^\mu =(y^+,y^-,\yT)$, with $y^\pm = (y^0\pm y^3)/\sqrt{2}$.
In all correlation functions, $y^+=0$ and we therefore do not
explicitly show the $y^+$ dependence.
The path ordered Wilson-line operator from the point $y$ to infinity
is defined as
\bea
\left[\infty^-,\yT;y^-,\yT \right] =
P\exp \left(-ig\int_{y^-}^\infty dz^- A^+(y^+,z^-,\yT)
\right).
\label{W1}
\eea
The specific choice of path in Eq. ({\ref{eq:P}) reflects the
FSI (ISI) of the active quark in an eikonal approximation.
The complex phase in Eq. ({\ref{eq:P}) is reversed under 
time-reversal
and therefore T-odd PDFs may exist \cite{collins},
which is why a nonzero Sivers asymmetry \cite{sivers} is possible.

Naively, the single spin asymmetry seems to be absent in light-cone 
gauge $A^+=0$, since the Wilson lines in Eq. (\ref{eq:P}) are
in the $x^-$ direction and therefore $\int dz^- A^+ =0$.
Without the phase factor any single spin asymmetry vanishes due to
time reversal invariance.

This apparent puzzle has been resolved in Ref. \cite{ji}, where it 
has been emphasized that a truly gauge invariant definition for
unintegrated parton densities requires closing the gauge link at
$x^-=\infty$, i.e. a fully gauge invariant version of Eq. 
(\ref{eq:P}) reads
\be
q(x,\kT,\sT) &=& \!\!\int \frac{dy^- d^2\yT }{16\pi^3}
e^{-ixp^+y^-+i\kT \cdot \yT } \label{eq:P2}
\\& &\!\!\times
\left\langle p \left|\bar{q}(y^-,\yT) \gamma^+
\left[y^-,\yT;\infty^-,\yT \right] 
\left[\infty^-,\yT,\infty^-,\0T\right]
\left[\infty^-,\0T;0^-,\0T\right] q(0)\right|p\right\rangle .
\nonumber
\ee
\begin{figure}
\unitlength1.cm
\begin{picture}(10,5)(5.3,21.5)
\includegraphics{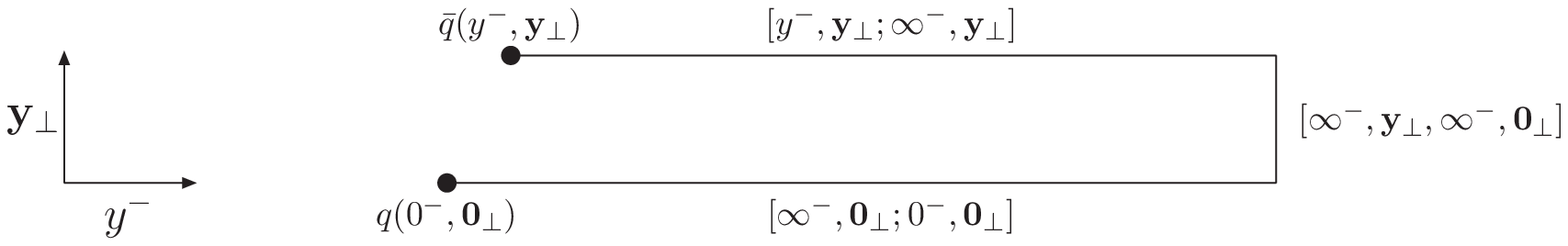}
\end{picture}
\caption{Illustration of the gauge links in Eq. (\ref{eq:P2}).}
\label{fig:staple}
\end{figure}  
In all commonly used gauges, except the light-cone 
gauge, the gauge link at $x^-=\infty$ is not expected to
contribute to the matrix element, since the gauge fields are 
expected to fall off rapidly enough at $\infty$. However, this is not
true in light-cone gauge and therefore it has been suggested in 
Ref. \cite{ji} that, in light-cone gauge, the entire single-spin 
asymmetry arises from the phase due to the gauge link at 
$x^-=\infty$
\be
q(x,\kT,\sT) &=& \int \frac{dy^- d^2\yT }{16\pi^3}
e^{-ixp^+y^-+i\kT \cdot \yT } \label{eq:P3}%\\& &\!\!\times
\left\langle p \left|\bar{q}(y^-,\yT) \left[\infty^-,\yT,\infty^-,\0T\right]
\gamma^+ q(0)\right|p\right\rangle .\quad \quad (A^+=0)
\ee
This implies for the average transverse momentum for quark flavor
$q$
\be
%\left\langle {\bf k}_{\perp q}\right\rangle &=&
\int d^2\kT q(x,\kT,\sT) \kT \label{k}
&=& -g\int \frac{dy^-}{4\pi}e^{-ixp^+y^-}
\left\langle p \left|\bar{q}(y^-,\0T) \gamma^+
{\bf A_\perp}(\infty^-,\0T) 
q(0)\right|p\right\rangle \quad \quad \quad (A^+=0)
\ee
A similar result holds for the unintegrated parton density relevant
for the Drell-Yan process, where the initial state interaction
provides the asymmetry
\bea
q_{past}(x,\kT,\sT) &=& \int \frac{dy^- d^2\yT }{16\pi^3}
e^{-ixp^+y^-+i\kT \cdot \yT } \label{eq:PDY}%\\& &\!\!\times
\left\langle p \left|\bar{q}(y^-,\yT) \left[-\infty^-,\yT,-\infty^-,\0T\right]
\gamma^+ q(0)\right|p\right\rangle \quad\quad (A^+=0)
\eea
and
\bea
\int d^2\kT q_{past}(x,\kT,\sT) \kT = 
-g\int \frac{dy^-}{4\pi}e^{-ixp^+y^-}
\left\langle p \left|\bar{q}(y^-,\0T) \gamma^+
{\bf A_\perp}(-\infty^-,\0T) 
q(0)\right|p\right\rangle  \quad\quad (A^+=0)
\label{kDY}
\eea
(Here and in the rest of this paper we will work in light-cone gauge
and therefore no longer emphasize explicitly that $A^+=0$).
The fact that these asymmetries hinge on the value of the transverse
gauge field at $x^-=\pm \infty$ makes the evaluation of these
matrix elements rather tricky.
Only a careful regularization prescription for the $k^+=0$ 
singularity of the gauge field propagator is capable to generate
the complex phase that is necessary for a non-vanishing SSA.
Therefore, the question arises
whether knowledge of the light-cone wave function for the nucleon
would (in principle) sufficient information  to calculate the SSA 
nonperturbatively, since the abovementioned prescription is
perturbatively defined.

One of the main purposes of this paper is to render Eqs. (\ref{k})
and (\ref{kDY}) into a more useful form.
For this purpose we first use the fact that the asymmetric part
of the unintegrated parton densities relevant for semi-inclusive DIS
and Drell-Yan are equal and opposite \cite{collins}
\be
\int d^2\kT q(x,\kT,\sT) \kT = - \int d^2\kT q_{past}(x,\kT,\sT) \kT
.
\ee
Therefore
\bea
%\left\langle {\bf k}_{\perp q}\right\rangle &=&
\bar{{\bf k}}_{\perp q}(x)&\equiv&
\int d^2\kT q(x,\kT,\sT) \kT \label{kAA}\\
&=& \frac{1}{2}\left[
\int d^2\kT q(x,\kT,\sT) \kT - \int d^2\kT 
q_{past}(x,\kT,\sT) \kT
\right]
\nonumber\\
&=&
-\frac{g}{2}\int \frac{dy^-}{4\pi}e^{-ixp^+y^-}
\left\langle p \left|\bar{q}(y^-,\0T) 
\left[{\bf A_\perp}(\infty^-,\0T) -{\bf A_\perp}(-\infty^-,\0T)\right]
\gamma^+ q(0)\right|p\right\rangle .\nonumber
\eea
For later applications, we also rewrite Eq. (\ref{kAA}) in the
more familiar color component notation
\bea
\bar{{k}}^i_{q}(x)
%\left\langle {\bf k}_{\perp q}\right\rangle
&=&
-\frac{g}{2}\int \frac{dy^-}{4\pi}e^{-ixp^+y^-}
\left\langle p \left|\bar{q}(y^-,\0T) \gamma^+ \frac{\lambda_a}{2}
q(0) { \alpha_a^i}(\0T)
\right|p\right\rangle
\label{kAAA}
\eea
where $\lambda_a$ are the Gell-Mann matrices and
\bea
{\bf \alpha_a^i}(\bT) \equiv
{\bf A_a^i}(\infty^-,\bT) -{\bf A_a^i}(-\infty^-,\bT).
\eea

Eqs. (\ref{kAA}) and (\ref{kAAA}) still involve the gauge field at 
$x^-=\pm\infty$.
In the following section, we will derive an operator relation
that relates ${\bf A_\perp}(\pm \infty,\yT)$ to degrees of freedom at
$-\infty < x^- < \infty$.

\section{Finiteness Conditions}
In a light-cone formulation of QCD ($x^+$ is time), not all 
components of the quark
and gluon field are dynamical, since they satisfy certain constraint
equations (Lagrangean contains no $x^+$ derivative for these degrees
of freedom). In particular, $A^-$ as well as the ``bad'' component
of the quark field $q_{bad}\equiv \frac{1}{2}
\gamma^+\gamma^- q$ are such constrained degrees of 
freedom in QCD. Upon eliminating both $A^-$ and $q_{bad}$ using
their constraint equations, one arrives at the canonical 
light-cone Hamiltonian for QCD in light-cone gauge $A^+=0$, 
which governs the $x^+$ evolution
\cite{reviews}
\bea
P^- = \int dx^- d^2\xT \left\{ F_+^\dagger \frac{1}{i\partial_-}F_+
+F_-^\dagger \frac{1}{i\partial_-}F_- + \frac{1}{2}\mbox{tr}\left[
-\tilde{J} \frac{1}{\partial_-^2}\tilde{J} + 
\left({\cal F}_{12}\right)^2\right] \right\}
\label{eq:P-}
\eea
where
\bea
{\cal F}_{12}&=&\partial_1A_2-\partial_2A_1 + ig\left[A_1,A_2\right]
\\
F_+&=&i\left[\partial_x+i\partial_y +ig\left(A_x+iA_y\right)
\right]u_- + \frac{m}{\sqrt{2}}u_+
\nonumber\\
F_-&=&i\left[\partial_x-i\partial_y +ig\left(A_x-iA_y\right)
\right]u_+ + \frac{m}{\sqrt{2}}u_-
\nonumber\\
\tilde{J}&=& 
\partial_-\partial_i A_i + J \nonumber\\
&=& \partial_-\partial_i A_i +
ig\left[A_i,\partial_-A_{i}\right] + g\left(
q_+q_+^\dagger + q_- q_-^\dagger\right)
\eea
In the above expression
%\bea
%A_z&=& \frac{1}{\sqrt{2}}\left(A_1-iA_2\right)\\
%A_{\bar{z}}&=& \frac{1}{\sqrt{2}}\left(A_1+iA_2\right)\nonumber\\
%\eea
%and 
$q_\pm$ form respectively the positive and negative chirality
components of the `good' component $q_{good}\equiv \frac{1}{2}
\gamma^-\gamma^+$ of the quark field.
A summation over quark flavors is implicit.

The requirement that $P^-$ acting on a hadron state is free of infrared $x^-=\pm \infty$ divergences  implies that each of the 
four terms in Eq. (\ref{eq:P-}) are free of such divergences.
This has a number of consequences.
In Ref. \cite{ladder}, it was investigated
what conditions on the hadron Fock state result from the requirement
that such divergences are absent in the first two terms. In this work we will focus on the third term. The condition
\bea
\int dx^-Tr\left[-\tilde{J}_\frac{1}{\partial_-^2}\tilde{J}\right] = \mbox{finite}
\eea
implies that \cite{zhang}
\bea
\int_{-\infty}^\infty dx^- \tilde{J}(x^-,\xT) = 0 \quad \quad 
\forall \xT
\label{J}
\eea
This (weak) condition, which should hold as a condition on all 
physical states, forms one of the crucial ingredients of this
investigation. Since the first term in $J$ is a total derivative,
Eq. (\ref{J}) implies (now expressed in terms of the more familiar
cartesian components $i=1,2$, which are implicitly summed when they
appear in pairs)
\bea
\partial^i \alpha^i(\xT)
= -\rho(\xT)\equiv -\int dx^- J(x^-,\xT),
\label{rho}
\eea
where
\bea
\alpha^i(\xT)&\equiv& A^i(\infty^-,\xT) - A^i(-\infty^-,\xT)
\nonumber\\
J (x^-,\xT) &=& ig\left[A^i,\partial_-A^i\right]
+ g \left( q_+q_+^\dagger + q_- q_-^\dagger\right).
\eea

The physical meaning of $\rho(\xT)$ is the total charge
(quarks plus gluons) along a line with fixed $\xT$.

For later use, we also express the above results using color 
components (instead of matrix notation)
\bea
\partial^i \alpha^i_a(\xT) &=& -\rho_a(\xT) \equiv
-\int dx^- J_a(x^-,\xT)
\label{nonabel1}\\
J_a(x^-,\xT) &=& -gf_{abc} A^i_b \partial_-A^i_c
+ g \sum_q \bar{q} \gamma^+\frac{\lambda_a}{2}q ,
\label{nonabel2}
\eea
where $\lambda_a$ are the Gell-Mann matrices. In QED, the
analogous conditions read
and
\bea
\partial^i \alpha^i(\xT) &=& -\rho(\xT) \equiv
-\int dx^- J(x^-,\xT)
\label{abel1}\\
J(x^-,\xT) &=& \sum_q e_q \bar{q}\gamma^+q .
\label{abel2}
\eea

An additional condition arises from the requirement that the 
$\mbox{tr}\left[\left({\cal F}_{12}\right)^2\right]$-term in the
Hamiltonian is convergent at $x^-=\pm \infty$:
the field $\perp$ strength tensor itself must vanish at 
$x^-=\pm \infty$
\bea
{\cal F}_{12}(\pm \infty^-,\xT)=0,
\eea
i.e. $A^j(\pm \infty^-,\xT)$ must be pure gauge \cite{ji}
\bea
A^j( +\infty^-,\xT) &=& -\frac{i}{g} V_+(\xT)\partial^j 
V^\dagger_+(\xT)\\
A^j( -\infty^-,\xT) &=& -\frac{i}{g} V_-(\xT)\partial^j 
V^\dagger_-(\xT).
\nonumber
\eea
This allows us to
gauge transform $A^j(-\infty^-,\xT)$ to zero while preserving
$A^+=0$. 
 Since in this gauge $A^j(+\infty^-,\xT) = 
$ is still pure gauge,
i.e. $A^j(\infty^-,\xT)=
-\frac{i}{g} U^\dagger(\xT) \partial^j U(\xT)$ with 
$U(\xT)= V_-^\dagger(\xT)V_+(\xT)$
we thus conclude that
\be
\alpha^i(\xT) =A^i(\infty^-,\xT)- A^i(-\infty^-,\xT)
\ee
must be (in this gauge) of the form
\be
\alpha^i(\xT) = -\frac{i}{g} U^\dagger(\xT) \partial^i U(\xT)
\label{gauge}.
\ee
Eq. (\ref{gauge}) together with Eq. (\ref{rho}) thus determine
$\alpha^i(\xT)$ uniquely (up to some trivial constants).

The above results have
a number of applications as we will discuss below.
First of all, Eqs. (\ref{nonabel1},\ref{abel1}) alert us again that in light-cone
gauge one must not assume a vanishing of the 
gauge fields at $x^-=\pm \infty$.
However, the most important application of Eqs. (\ref{nonabel1},\ref{abel1})
lies in the fact that it allows us to reexpress 
$\alpha^i(\xT)=A^i(\infty^-,\xT) - A^i(-\infty^-,\xT)$ in terms of 
other degrees of freedom.
The interesting aspect about this observations is the fact that
$\alpha^i(\xT)$ also appears in the correlation function 
(\ref{kAA}) for the average transverse momentum. 
In the rest of this paper we will discuss the implication of this
fundamental result. 
\section{Average $\perp$ momentum in the abelian case}
Before proceeding to QCD, we will first discuss an abelian theory,
where the nucleon contains quarks of different flavor $q$ with
charges $e_q$ respectively as well as photons. For simplicity, we will
consider here only the Sivers assymmetry averaged over all $x$
\be
\left\langle{\bf k}_{\perp q} \right\rangle &=&
\int dx \int d^2\kT q(x,\kT,\sT)\kT = \int dx \bar{{\bf k}}_{\perp q}(x).
\ee
Considering only the $x$-averaged asymmetry not only
helps keep the resulting expressions simpler, but may also help
to cancel possible divergences from endpoint singularities \cite{co3}.

In QED, the requirement that $\alpha^i(\xT)$ is pure gauge can be 
rewritten as
\bea
\alpha^i(\xT) = -\partial^i \phi (\xT),
\eea
where $\phi(\xT)$ is some scalar function, which can be determined by
solving the 2-dimensional Poisson equation. This yields
\bea
\alpha^i(\xT) 
= -\int \frac{d^2\yT}{2\pi} \frac{x^i-y^i}{\left|\xT-\yT\right|^2}
\rho(\yT).
\label{soliton}
\eea
What we have found is an operator condition (\ref{soliton}) that
determines the transverse gauge field at $x^-=\pm \infty$
in terms of the charge density at $-\infty<x^-<\infty$.
This condition needs to be satisfied for states in order to have an 
infrared finite energy. Upon inserting this result into the
expression for $\left\langle {\bf k}_{\perp q}\right\rangle$ we find
\bea
\left\langle{\bf k}_{\perp q} \right\rangle = -\frac{e_q}{4p^+}
\int \frac{d^2\yT}{2\pi} \frac{\yT}{\left|\yT\right|^2}
\left\langle p \left|
\bar{q}(0)\gamma^+ q(0) \rho(\yT) \right| p \right\rangle
\label{qed}
\eea
where
\bea
\rho(\yT) = \sum_{q^\prime} e_{q^\prime} \int dy^- \bar{q}^\prime(y)
\gamma^+ q^\prime(y)
\eea
is the charge density (integrated over $x^-$) from all flavors
(actually, for symmetry reasons, $q^\prime =q$ does not contribute
in Eq. (\ref{qed}).

Eq. (\ref{qed}) is gauge invariant and provides a regularized
expression for the Sivers asymmetry that depends on the light-cone
wave function of the target only. Although this should be clear from
our derivation, it should be emphasized that the final state 
interactions are included in Eq. (\ref{qed}), but the gauge fields
that give rise to the final state interactions have been
reexpressed in terms of charge density correlations inside the target.
The relevant density-density correlations are correlations in the 
transverse plane of the charge density integrated along $x^-$.
The kernel in the correlation function is the Lorentz-boosted
Coulomb force integrated along $x^-$ as well.
Such a result should not be surprising since this is simply
the Coulomb force from the spectators acting on the escaping quark.
In fact, for the specific example of the scalar diquark model this
result was already obtained in Ref. \cite{me:hwang}.

As a corollary, we should also note that if one sums the mean 
transverse momentum over all quark flavors one gets zero
\bea
\sum_q \left\langle {\bf k}_{\perp q}\right\rangle =0.
\label{sum0}
\eea
This result follows from the fact that the integration kernel 
in Eq. (\ref{soliton}) is odd under $\xT \leftrightarrow \yT$.

To summarize this section, what we have found is that the mean
transverse momentum for quarks of flavor $q$ can be related to
the correlations between quarks of flavor $q$ and all other quarks
in the transverse plane. This is not surprising since the final state
interaction is the Lorentz boosted Coulomb interaction and the
correlations describe the Coulomb field from the spectators acting
on the active quark.

\section{Average $\perp$ momentum in QCD}
In QCD, the condition that the gauge field at $\pm \infty$ is
pure gauge is nonlinear, which prevents us from writing down
closed form solutions to the finiteness conditions. This can be 
seen as follows. If one writes 
\be
U(\xT) = e^{-ig\phi_a(\xT)\frac{\lambda_a}{2}}
\ee
then, to lowest order in $\phi_a$ (see also Appendix B)
one finds the QED-like condition
\be
\Delta \phi_a (\xT) = -\rho_a(\xT),
\label{poisson}
\ee
yielding
\bea
\alpha_a^i(\xT) = -\partial^i \phi_a
= -\int \frac{d^2\yT}{2\pi} \frac{x^i-y^i}{\left|\xT-\yT\right|^2}
\rho_a(\yT).
\label{soliton2}
\eea
Of course, there are nonabelian corrections to Eq. (\ref{poisson})
and therefore, unlike in QED, Eq. (\ref{soliton2}) is not an exact
solution to the finiteness conditions. However, since we were unable
to find an exact operator solution, we will proceed using Eq. 
(\ref{soliton2}).

Upon inserting (\ref{soliton}) into Eq. (\ref{kAAA}) one obtains
the nonabelian version of Eq. (\ref{qed})
\bea
\left\langle k^i_q \right\rangle = -\frac{g}{4p^+}
\int \frac{d^2\yT}{2\pi} \frac{y^i}{\left|\yT\right|^2}
\left\langle p \left|
\bar{q}(0)\gamma^+ \frac{\lambda_a}{2}q(0) \rho_a(\yT) 
\right| p \right\rangle
\label{qcd}.
\eea
The physical interpretation of this result is that the average
transverse momentum of quarks of flavor $q$ can be related to 
correlations on the transverse plane. The specific correlations that
appear in Eq. (\ref{qcd}) reflect a Coulomb interaction between the
active quark and the spectators. Here a Coulomb interaction appears
because we have solved the finiteness constraints in QCD only to first
order. Eq. (\ref{qcd}) is thus equivalent to treating the FSI in
lowest order in perturbation theory \cite{hwang,feng,me:hwang,regen}.
Note also that the resulting correlation functions are very similar to the correlation functions that have been used to describe the small-$x$ gluon distributions in nuclei \cite{raju}.

To first order, what we also find is that the average transverse
momentum due to the FSI of all constituents (quarks + gluons)
added together vanishes for symmetry reasons. It is not clear if this
happens beyond lowest order.

Eq. (\ref{qcd}) may be useful for several reasons. While the original 
expression for the Sivers asymmetry involved a gauge link, which
made a parton model interpretation difficult, Eq. (\ref{eq:36})
does have an immediate parton model interpretation in terms of 
color-flavor correlations in the transverse plane. This may be useful
in correlating experimental data with our understanding of the
nucleon structure. Another use of Eq. (\ref{qcd}) is that it can be
directly calculated from the light-cone wave functions of the nucleon.
Of course, we need to keep in mind that (unlike the QED case)
Eq. (\ref{qcd}) is only approximation, but we still believe that this
result provides a step towards linking the Sivers asymmetry with 
other features of hadron structure. 
Finally, we would like to emphasize that Eq. (\ref{qcd}) suggests
interesting connections between the distribution of partons in
impact parameter ($\xT$) and the sign of the transverse SSA
\cite{me}. For example, in a simple quark model, such as the bag model
\cite{feng}, the color part of the matrix element in Eq. 
(\ref{qcd}) would be negative (attraction). If the transverse
distribution of $j^+_q$ is transversely shifted relative
to the spectators \cite{ijmpa} then the resulting average transverse
momentum has the opposite sign of the sign of the transverse 
distortion in impact parameter space.

\section{Summary}
We have studied the average transverse momentum of gauge invariant
Quark distributions for a transversely polarized target in light-cone gauge.
The Wilson line is along the light-cone to infinity to incorporate the
Final state interactions in semi-inclusive DIS.
In light-cone gauge, the Wilson-line phase factor receives its only nonzero
contribution from the gauge field at $x^-=\pm \infty$. 
In a naive Fock space expansion the gauge field at $x^-=\pm \infty$
is usually implicitly set to zero, thus making a correct treatment
of single-spin asymmetries rather difficult (except in perturbation
theory, where one can carefully regularize the fields at 
$x^-=\pm \infty$ ``by hand''.

We have also studied conditions for the infrared ($x^-=\pm \infty$) 
convergence of the
light-cone Hamiltonian for gauge theories and derived operator
conditions that need to be satisfied in order for the light-cone to
be free of infrared divergences arising from otherwise ill-defined
operators $\frac{1}{i\partial_-}$. This operator condition relates
the transverse component of the gauge field $A^i_a(\pm \infty^-,\xT)$
to the color density $\rho_a(\xT)$ integrated over all $x^-$
 
Fortuitously, the same kind of operators 
that governs the average transverse momentum in gauge invariant
quark distributions also appears in the finiteness conditions for
the light-cone Hamiltonian. We are thus able to eliminate 
$A^i_a(\pm \infty^-,\xT)$ in the average transverse momentum
in favor of other, less infrared singular, degrees of freedom.
In QED we can solve the operator condition arising from finiteness
conditions exactly and we are able to express the average transverse
momentum in terms of charge density correlations in the transverse
plane. In QCD we were only able to solve the operator condition to
first order in the color charge density and there we find a 
similar result as in QED, namely that the average transverse 
momentum can be related to transverse correlations between the
active quark and the spectators.

Single spin asymmetries do not have a simple parton model
(or light-cone Fock space) interpretation. The main significance
of our results is that we have found relations that allow to
relate the average transverse momentum to operators that do have
a parton interpretation (in QED exactly, in QCD approximately).
One immediate application of these results is that it allows to
evaluate the average transverse momentum of the quarks directly
from the nucleon wave function in light-cone quark models.

Several extensions of this work are conceivable. First it would
be desirable to derive an exact solution (at least in terms of an
expansion) for the finiteness conditions in QCD, so that one can
study the effects of higher order terms that we have omitted.
Secondly, it would be interesting to see if one can translate
the results from the work into lattice language (Euclidean as
well as transverse lattice) with the goal of being able to
compute the average transverse momentum nonperturbatively within
these frameworks.

This work was supported by the Department of Energy 
(DE-FG03-96ER40965). I would like to thank J. Collins and X. Ji
for useful discussions and comments.
\appendix
\section{Alternative Derivation fo the Finiteness Condition}
In this appendix we would like to present an alternative derivation 
of Eq. (\ref{nonabel2}) that does not require a light-cone 
Hamiltonian. We start from the $+$ component of the QCD
equations of motion in light-cone gauge $A^+=0$
\be
D_\mu {\cal F}^{\mu +}_a \equiv \partial_- {\cal F}^{-+}_a
+ \partial^i {\cal F}^{i+}_a -g f_{abc} A^i_{b} {\cal F}^{i+}_c = 
j^+_a
\label{field}
\ee
where
\bea
{\cal F}^{-+}_a &=& \partial^-A^+_a -\partial^+A^-_a -gf_{abc}
A^-_bA^+_c = -\partial_-A^-_a\\
{\cal F}^{i+}_a &=& \partial^iA^+_a -\partial^+A^i_a -gf_{abc}
A^i_bA^+_c = -\partial_-A^i_a\nonumber 
\eea
and the fermion color density is given
\be
j^+_a \equiv g\sum_q \bar{q}\gamma^+ \frac{\lambda_a}{2} q 
\ee  
is the fermion color density.
Written out in components the Field equations (\ref{field}) thus read
\bea
-\partial_-^2 A^-_a -\partial_-\partial^i A^i_a - gf_{abc} 
A^i_{b} {\cal F}^{i+}_c = j^+_a
\label{field2}
\eea
Integrating Eq. (\ref{field2}) over $x^-$, while making use of the
condition that the field strength tensor $F^{+-}_a$ at 
$x^- = \pm \infty$ vanishes, thus yields
\bea
\label{Arho}
\partial^i A^i_a(\infty^-,\xT) - \partial^iA^i_a(-\infty^-,\xT)
&=& -\int_{-\infty}^\infty dx^- 
\left[ j^+_a - g f_{abc}A^i_{b} \partial_- A^i_c\right] 
\\
&=& -\int _{-\infty}^\infty dx^- J_a (x^-,\xT) = - \rho_a(\xT)
\equiv \rho(\xT).
\nonumber
\eea
\section{Higher Order Color Correlations}
We want to solve the finiteness condition
\bea
\partial^i \alpha^i_a(\xT) = -\rho_a (\xT)
\label{Brho}
\eea
subject to the constraint that $\alpha^i_a(\xT)$ is pure gauge, i.e.
\bea
\alpha^i(\xT) = -\frac{i}{g} U^\dagger \partial^i U.
\eea
In order to satisfy the second condition, one can make the ansatz
\be
U(\xT) = \exp\left(-ig\phi_a(\xT) \frac{\lambda_a}{2}\right).
\ee
Upon inserting this ansatz into Eq. (\ref{Arho}) one finds
\be
\alpha^i(\xT) = - \partial^i\phi(\xT) - \frac{ig}{2} \left[\phi(\xT),\partial^i\phi(\xT)
\right] + \frac{g^2}{12} \left[\phi(\xT),\left[\phi(\xT),\partial^i\phi(\xT)\right]
\right]+ ...
\ee
i.e. in component notation 
\be
\alpha_a^i(\xT) = - \partial^i \phi_a(\xT) + \frac{g}{2} f_{abc} \phi_b(\xT)
\partial^i \phi_c(\xT)-\frac{g^2}{12} f_{abc} \phi_b(\xT) f_{cde} \phi_d(\xT)
\partial^i\phi_e(\xT)+...
\label{series0}
\ee
and therefore
\be
\partial^i \alpha_a^i(\xT)&=& - \partial^i\partial^i\phi_a(\xT)
+ \frac{g}{2} f_{abc} \phi_b(\xT)
\partial^i\partial^i \phi_c(\xT) \label{series}
\\
& &-\frac{g^2}{12} f_{abc} f_{cde}\left[ \phi_b(\xT)  \phi_d(\xT)
\partial^i\partial^i\phi_e(\xT)+  \partial^i\phi_b(\xT)  \phi_d(\xT)
\partial^i\phi_e(\xT)\right]
+... \nonumber\\
&\stackrel{!}{=}&-\rho_a(\xT) \nonumber
\ee
One may attempt to solve Eq. (\ref{series}) by making a formal expansion 
in powers or $\rho_a(\xT)$
\be
\phi_a(\xT) = \phi^{(0)}_a(\xT) + \phi^{(1)}_a(\xT)+...,
\ee
yielding
\be
\partial^i\partial^i\phi^{(0)}_a(\xT) &=& \rho_a(\xT)\label{series2}\\
\partial^i\partial^i\phi^{(1)}_a(\xT) &=& \frac{g}{2} f_{abc} \phi^{(0)}_b(\xT)
\partial^i\partial^i \phi^{(0)}_c(\xT)  = \frac{g}{2} f_{abc} \phi^{(0)}_b(\xT)
\rho_c(\xT).
\nonumber\\
&e.t.c.& \nonumber
\ee
Eq. (\ref{soliton2}) is then obtained by keeping only the lowest order term in
Eqs. (\ref{series0}) and (\ref{series2}).
\bibliography{zero3.bbl}% Produces the bibliography via BibTeX.

\begin{thebibliography}{3}
\bibitem{lambda} 
P.J. Mulders, lectures at PRAHA 2001, 
hep-ph/0112225; D. Boer and J. Qiu, Phys.\ Rev.\
D\ {\bf 65}, 034008 (2002). 
\bibitem{collinsold} J.C. Collins, Phys. Lett. B {\bf 396}, 161 
(1993).
\bibitem{sivers} D.W. Sivers, Phys.\ Rev.\ D {\bf 43}, 261 (1991).
\bibitem{hwang} S.J. Brodksy, D.S. Hwang, and I. Schmidt,
Phys.\ Lett.\ B {\bf 530}, 99 (2002); S.J. Brodsky et al., Phys.\
Rev.\ {\bf D65}, 114025 (2002).
\bibitem{collins} J.C. Collins, Phys. Lett. B {\bf 536}, 43 (2002).
\bibitem{ji} X. Ji and F. Yuan, Phys. Lett. B {\bf 543}, 66 (2002);
A. Belitsky, X. Ji, and F. Yuan, Nucl. Phys. B {\bf 656}, 165 (2003).
\bibitem{boer} D. Boer, P.J. Mulders, and F. Pijlman, hep-ph/0303034. 
\bibitem{reviews} M. Burkardt, Adv. Nucl. Phys. {\bf 23}, 1 (1996);
S.J. Brodsky, H.C. Pauli and S.S. Pinsky, Phys Rept. {\bf 301}, 299
(1998).
\bibitem{ladder} F. Antonuccio, S.J. Brodsky, and S. Dalley,
Phys.\ Lett. {\bf B412}, 104 (1997).
\bibitem{zhang} W.M. Zhang, in ``Light-Front Quantization and
Non-Perturbative QCD'', Eds. J. Vary and F. W\"olz, (IITAP, 1997),
p. 141; M. Burkardt, in ``New Directions in Quantum Chromodynamics'',
AIP Conf. Proc. {\bf 494}, p. 239 (1999); hep-th/9908195.
\bibitem{feng} F. Yuan, Phys. Lett. {\bf B575}, 45 (2003).
\bibitem{co3} J. Collins, Acta Phys. Polon. {\bf B34}, 3103 (2003).
\bibitem{me:hwang} M. Burkardt and D.S. Hwang, hep-ph/0309072.
\bibitem{regen} A. Bacchetta, A. Sch\"afer, and J.-J. Yang,
hep-ph0309246.
\bibitem{me} M. Burkardt, Phys.\ Rev.\ D {\bf 66}, 114005 (2002);
M. Burkardt, hep-ph/0302144.
\bibitem{raju} L. McLerran and R. Venugopalan, Phys.\ Rev.\ D
{\bf 59}, 094002 (1999).
\bibitem{ijmpa} M. Burkardt, Int.\ J.\ Mod.\ Phys.\ 
A {\bf 18}, 173 (2003); X. Ji, hep-ph/0304037.
\end{thebibliography}
\end{document}